# The Effect of Information Type on Human Cognitive Augmentation

Ron Fulbright[1], Samuel McGaha[2]
University of South Carolina Upstate
800 University Way, Spartanburg, SC USA 29303
[1]fulbrigh@uscupstate.edu
[2]sm30@email.uscupstate.edu

**Abstract.** When performing a task alone, humans achieve a certain level of performance. When humans are assisted by a tool or automation to perform the same task, performance is enhanced—augmented. Recently developed cognitive systems are able to perform cognitive processing at or above the level of a human in some domains. When humans work collaboratively with such "cogs" in a human/cog ensemble, we expect augmentation of cognitive processing to be evident and measurable. This paper shows the degree of cognitive augmentation depends on the nature of the information the cog contributes to the ensemble. Results of an experiment are reported showing conceptual information is the most effective type of information resulting in increases in cognitive accuracy, cognitive precision, and cognitive power.

## 1 Introduction

Recent developments, most notably in the fields of unsupervised deep learning, have produced systems capable of outperforming human experts in many domains. Computers have outplayed human experts in various games such as card games, Checkers, and Chess for several years and within the last decade have conquered human champions in *Jeopardy!* and Go (Ferrucci, 2012; Silver, et al., 2016; DeepMind, 2018). Going far beyond gameplaying, systems now diagnose cancer, childhood depression, dementia, heart attacks, achieve reading comprehension, and discover new patterns in mathematics better than human experts (Wehner, 2019; Lavars, 2019; Towers-Clark, 2019; Gregory, 2019). These systems are not artificially intelligent, yet they mimic, perform, or replace parts of human-level thinking. Systems like this are called *cognitive systems*, or "cogs" for short (Wladawsky-Berger, 2015; Gil, 2019; Kelly & Hamm, 2013).

Cogs like these are assistive tools used by humans in a collaborative engagement called a *huma/cog ensemble*. Aggregate cognitive processing of a human/cog ensemble is therefore a mixture of artificial and biological thinking and exceeds the cognitive processing of a human acting alone. Using cogs, augmented humans outperform unassisted humans therefore we say the human is cognitively augmented. If cognitive performance is enhanced, we should be able to measure it. To do so requires us to measure either the information itself, the cognition, or the results of the cognition. Neither of these is an easy task yet. However, theoretical and practical work is progressing.

This paper presents the results of an experiment designed to investigate hypothesis H1, shown below, that the degree of cognitive augmentation achieved in a human/cog ensemble is dependent on the nature of information supplied to the human by the cog.

> **H1:** The degree of cognitive augmentation achieved by humans working together on a task in collaboration with cognitive systems is dependent on the nature of information contributed by the cognitive system.

To investigate the hypothesis, we performed an experiment asking humans to solve several non-trivial puzzles. To simulate different contributions of a cognitive system, some humans were given no assistive information whereas others were given assistive information of two different types: *conceptual* and *policy/principle*. Results showed both types of assistive information improved performance, but conceptual information had greater impact on cognitive performance than policy/principle information. Furthermore, we were able to calculate cognitive augmentation by calculating increases in cognitive accuracy and cognitive precision.

## 2 Literature and Previous Work

**2.1 Measuring Cognitive Augmentation**

We can view data, information, knowledge, and wisdom (DIKW) as a hierarchy based on value as shown in Fig 1. (Ackoff, 19989). Data is obtained by sensing disturbances in the environment, information is processed data, knowledge is processed information, and wisdom is processed knowledge. Each level is of a higher value than the level below it because of the processing involved and the utility of information stock at that level.

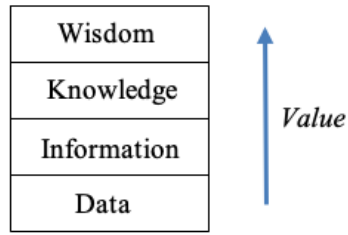

**Fig. 1.** The DIKW Hierarchy.

Processing at each of the level in the DIKW can be modeled as a cognitive process transforming data, information, or knowledge, generically referred to as *information stock*, to a higher-valued form as depicted in Fig 2 where the transformation of the information stock is accomplished by the expenditure of a certain amount of *cognitive work* (W) (Fulbright, 2020).

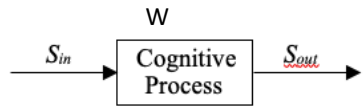

**Fig. 2.** A cognitive process as a transformation of information stock.

To illustrate how cognitive processing increases the value of information stock, consider a temperature sensor. The electrical conductivity of two different metals in a thermocouple is affected by the temperature of the environment in which the thermocouple is placed causing a detectable voltage potential. Detecting the voltage represents *data*—a direct sensing of a physical disturbance. To convert this reading to temperature a calculation must be performed. The calculation is a cognitive process combining the data sensed from the environment with information obtained from an engineering units reference table, and the knowledge of how to calculate the formula. The result of this cognitive process is *degrees* and represents new information of a higher value than the data input into the cognitive process. Similarly, information is processed into knowledge and knowledge is processed into wisdom by additional cognitive processing.

In a human/cog ensemble (a collaborative team), cognitive processing of the entire ensemble is a mixture of human cognitive processing and artificial cognitive processing ($W^* = W_H + W_C$) as depicted in Fig. 3 (Fulbright, 2020; Fulbright & Walters, 2020; Fulbright, 2020a).

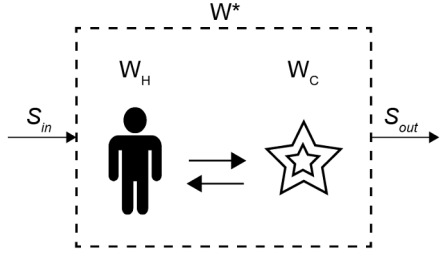

**Fig. 3.** A Human/Cog ensemble performing a cognitive process.

In earlier work, we have proposed several methods of measuring and calculating cognitive work and the degree of cognitive augmentation achieved in a human/cog ensemble (Fulbright, 2017; Fulbright, 2018; Fulbright, 2019; Fulbright, 2020). A way of measuring the amount of cognitive work done by a cognitive process is to compare the value of the information stock before and after the processing as shown in Eq. (1) where the value of the information stock is evaluated by the value function, $\psi$.

$$W = |\psi(S_{out}) - \psi(S_{in})| \tag{1}$$

Eq. (1) therefore, focuses on the transformation effected by the cognitive process. One way to measure cognitive augmentation is to calculate a quantity called *cognitive power* as shown in Eq. (2) where W represents an amount of cognitive work performed by one or more cognitive processes and $t$ is the time required to perform W.

$$P = \frac{W}{t} \tag{2}$$

In general, cognitive power increases as the amount of cognitive work increases or the amount of time decreases. In a human/cog ensemble, contributions by either the human or the cog can result in either.

Another way to measure cognitive augmentation is to measure the increase in *cognitive accuracy* and/or *cognitive precision* of an augmented human. Cognitive accuracy is a measure of the ability to produce the correct, or preferred, output. Cognitive precision is a measure of the ability to produce *only* the correct or preferred output as depicted in Fig. 4 where the oval represents the correct or preferred output.

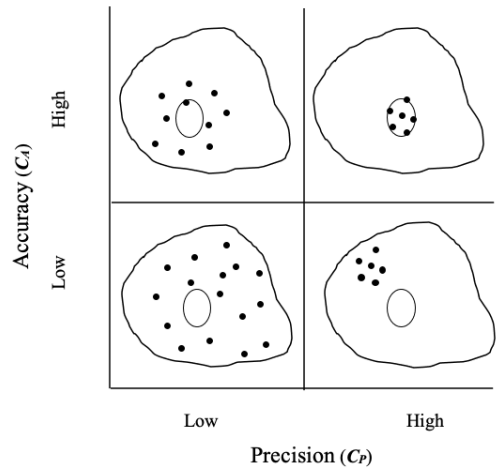

**Fig. 4.** Cognitive Accuracy and Cognitive Precision.

The goal, of course, is to achieve high accuracy *and* high precision (upper right quadrant of Fig. 4). Using a chosen accuracy and precision performance metric (*x* and *y*), comparing the performance of a human working alone (*x*) to a human working in partnership with a cog in a human/cog ensemble (*x'*) will calculate any change in cognitive accuracy and cognitive precision as shown in Eq. (3).

$$\Delta C_A = \frac{x-x'}{x} \qquad \Delta C_P = \frac{y-y'}{y} \qquad (3)$$

For example, a human working alone might produce the correct result 4 times out of 10. If the same human, working in partnership with a cog, produces the correct result 8 times out of 10, then the cognitive accuracy has increased by two-fold, a 100% increase.

Human performance is *augmented* by partnering with cogs and is superior to humans acting alone. However, not all human/cog ensembles result in the same level of cognitive augmentation. Different Levels of Cognitive Augmentation have been defined ranging from no augmentation at all (all human thinking) to fully artificial intelligence (no human thinking) as shown in Fig. 5 (Fulbright, 2020; Fulbright & Walters, 2020; Fulbright, 2020A).

**Level 0**: No Augmentation
human performs all cogntiive processing

**Level 1**: Assistive Tools
abacus, calculators, software, etc.

**Level 2**: Low-Level Cognition
pattern recognition, classification, speech
human makes all high-level decisions

**Level 3**: High-Level Cognition
concept understanding, critique,
conversational natural language

**Level 4**: Creative Autonomy
human-inspired, unsupervised synthesis

**Level 5**: Artificial Intelligence
no human cognitive processing

**Fig. 5**. Levels of Cognitive Augmentation.

## 2.2 Types of Information
Earlier, we characterized various types of information (data, information, knowledge, and wisdom) based on processing and the utility value of the information at the various levels. However, DIKW is not the only way to characterize information. Hertz and Rubenstein identified six types of information as shown in Fig. 6 (Hertz & Rubenstein, 1953; LISBON, 2014; Indeed, 2021).

- **Conceptual**     ideas, theories, hypotheses, etc.
- **Procedural**     method of how to do something
- **Policy**         laws, rules, guidelines, theorem, etc.
- **Stimulatory**    information causing activity
- **Empirical**      information through sensing, observation, experiment, etc.
- **Directive**      information provided to achieve a particular outcome
- **Fact**           a statement of data asserted with certainty

**Fig. 6** Hertz and Rubenstein's Six Types of Information.

Robert Horn, the developer of Information Mapping™, identified seven types of information as shown in Fig. 7 (Horn, 1966; Horn, 1989; Horn, 1993).

- **Procedure**        a set of specific steps and decisions to be made
- **Process**          a series of events or phases taking place over time
- **Concept**          a group or class of objects, conditions, events, ideas etc.
- **Structure**        physical structure divided into parts by boundaries
- **Classification**   division of items into categories using one or more factors
- **Principle**        rule, policy, guideline, theorem, axiom, postulate, etc.
- **Fact**             a statement of data asserted with certainty

**Fig. 7** Horne's Seven Types of Information.

Even though these two sources use different names and words, the categories of information types defined are very similar. In our experiment, we chose to use *conceptual* and *policy (*also called *principle)* information. Examples of conceptual information include definitions, examples, and counter examples. Examples of principle information include guidelines, rules, goals, and objectives (Thomas, 2004).

## 3    The Experiment

Participants were asked to solve four different puzzles listed below and shown in Fig. 8.

- **Task 1:** "Square"      (3-row math puzzle)
- **Task 2:** "X puzzle"    (diagonal math puzzle)
- **Task 3:** "4 X 4"       (4-row sequence puzzle)
- **Task 4:** "Message"     (6-word decryption puzzle)

**Fig. 8** Four puzzles participants were asked to solve.

The puzzles were presented to the participants one at a time with the participant allowed to continue to the next puzzle only upon successful completion of the current puzzle. Two of the four puzzles involved basic mathematical functions (addition, subtraction, multiplication). One puzzle involved recognizing a pattern in a sequence of numbers. One puzzle involved solving decoding a simple substitution cyber. Each puzzle involved non-trivial kinds of cognition but was simple enough to be solved by anyone with grade-school education and knowledge.

To investigate the effect of different types of information, some participants were presented with a hint along with the puzzle. Approximately 1/3 of the participants were given no hint (the "normal" group) and served as the control group. Approximately 1/3 of the participants were given a hint in the form of conceptual information ( the "concept" group). The conceptual hint was an example of a completed puzzle shown to the participants. The remaining 1/3 of the participants were given a hint in the form of principle/policy information (the "policy" group). The policy/principle hint for each puzzle involved a guideline or rule as shown below:

- **Square**         "Each row is a different mathematical operation."
- **X puzzle**       "The middle box and the empty box combine to equal the third box."
- **4 X 4**          "Each row is based on a specific number. One row is a combination of the other three rows."
- **Message**        "Each number is tied to a specific letter in the English alphabet."

To take part in the experiment, participants downloaded a computer program presenting each of the four puzzles and the assistive information (if any). Participants were given up to one hour to complete the puzzles. If, after an hour, all puzzles were not solved the attempt was counted as a failure. Participants were allowed to submit an attempted solution to a puzzle and then receive a message whether the solution was correct. If incorrect, the participant was allowed to repeat and submit another solution. Attempted solutions were limited to 25. If after 25 attempts the puzzles were not solved, the attempt was listed as a failure. Performance of the participants was assessed in several ways:

- Failure Percentage (inability to solve a puzzle)
- Total Overall Time (total time taken working on the puzzles)
- Average Attempts Per Puzzle
- Longest Individual Time per Puzzle
- Shortest Individual Time per Puzzle
- Highest Individual Number of Attempts per Puzzle
- Lowest Individual Number of Attempts per Puzzle

## 4  The Results

**4.1 Failure Percentage**
During the testing phase, some participants failed to complete the puzzles within 25 attempts or one hour of time. Participants receiving conceptual information as a hint (the "concept" group) had the least number of failures whereas those receiving no information at all (the "normal" group) had the most failures as seen in Fig. 9. Success of the "concept" group was three times better than the "normal group."

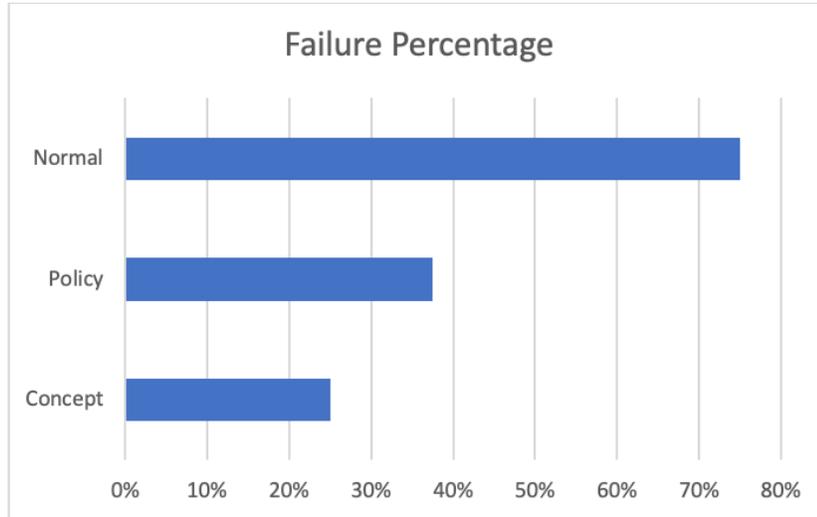

**Fig. 9** Failure Percentage for Different Types of Information.

Failure percentage ($F$) is a measure of *accuracy*. A failure percentage of 100% would mean a complete lack of accuracy and a failure percentage of 0% would mean perfect accuracy. Using, Eq. (3), the decrease in failure percentage for each type of information can be calculated showing conceptual information has the greatest impact on failure percentage:

$$\Delta F_{Policy} = \frac{75\% - 37\%}{75\%} = 51\% \qquad (4)$$

$$\Delta F_{Conceptual} = \frac{75\% - 25\%}{75\%} = 67\% \qquad (5)$$

The inverse of failure percentage is also a measure of cognitive accuracy. Participants receiving conceptual information as a hint were successful 75% of the time. Participants receiving policy/principle information were successful 63% of the time. Participants receiving no assistive information were successful only 25% of the time. Therefore, when compared to no information, policy/principle information increased cognitive accuracy by 60% ($\Delta C_A = 60\%$), a 1.7 fold increase, and conceptual information increased cognitive accuracy by 200% ($\Delta C_A = 200\%$), a three-fold increase.

### 4.2 Total Overall Time

The total overall time for a group of participants is the sum of all times spent by participants in the group, measured in seconds. Participants receiving conceptual information as a hint had the shortest overall time (the "concept" group) whereas those receiving no information at all (the "normal" group) had the longest overall time as seen in Fig. 10. The "concept" group spent less than half the amount of time the "normal" group did.

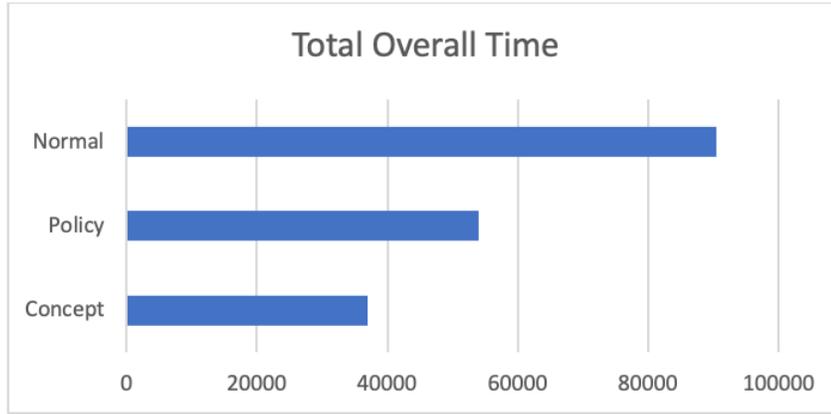

**Fig. 10** Total Overall Time (*in seconds*).

By calculating the reduction in time, we see conceptual information had the greatest impact on time spent on the puzzles.

$$\Delta T_{Policy} = \frac{90,000-55,000}{90,000} = 39\% \tag{6}$$

$$\Delta T_{Conceptual} = \frac{90,000-35,000}{90,000} = 61\% \tag{7}$$

Before any cognitive processing was done, the four puzzles were in the unsolved state with a certain amount of value associated. After successfully solving the four puzzles, they were in the solved state at an increased value. Therefore, according to Eq. (1), a nonzero amount of cognitive work was performed by the participants (W > 0). Therefore, using Eq. (2), cognitive power can be calculated for each type of information as shown in Eq. (8).

$$P_{Normal} = \frac{W}{90,000} < P_{Policy} = \frac{W}{55,000} < P_{Conceptual} = \frac{W}{35,000} \tag{8}$$

Cognitive augmentation by virtue of conceptual information yielded a cognitive power more than 2.5 times greater than no information (no augmentation) and more than 1.5 times that of policy/principle information.

### 4.3 Average Attempts per Puzzle
Participants were allowed to attempt each puzzle multiple times (up to 25 times). The number of attempts for each group is the average of the number of attempts for each participant in a group for each puzzle. Participants receiving conceptual information as a hint (the "concept" group) had the fewest number of attempts for each puzzle whereas those receiving no information at all (the "normal" group) had the greatest number of attempts for each puzzle as seen in Fig. 11. The "normal" group had three times the number of attempts over the "concept" group.

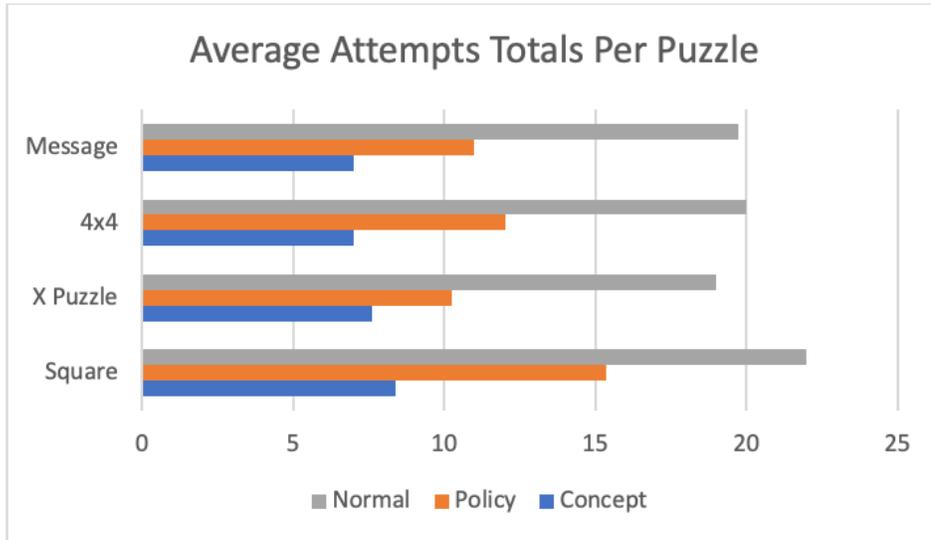

**Fig. 11** Average Attempts Per Puzzle.

The number of attempts per puzzle is a measure of *precision*. Correct solution of a puzzle on the first try would represent maximal cognitive precision with cognitive precision decreasing as the number of incorrect attempts increases. For each puzzle, comparing the impact of policy/principle information and conceptual information against no information yields a cognitive augmentation attributed to conceptual information increased cognitive precision of 63% -65%.

**Message Puzzle:** $\Delta C_P(policy) = \frac{19-11}{19} = 42\%$ $\quad \Delta C_P(Conceptual) = \frac{19-7}{19} = 63\%$

**4x4 Puzzle:** $\Delta C_P(policy) = \frac{20-12}{20} = 40\%$ $\quad \Delta C_P(Conceptual) = \frac{20-7}{20} = 65\%$

**X Puzzle:** $\Delta C_P(policy) = \frac{20-12}{20} = 40\%$ $\quad \Delta C_P(Conceptual) = \frac{20-7}{20} = 65\%$

**Square Puzzle:** $\Delta C_P(policy) = \frac{22-16}{22} = 27\%$ $\quad \Delta C_P(Conceptual) = \frac{22-8}{22} = 64\%$

### 4.4 Longest and Shortest Individual Time per Puzzle

Participants were allowed to spend as much time as they wished on each puzzle. Since each puzzle required different types and kinds of cognitive effort to complete, time (measured in seconds) spent on each puzzle varied:

| | |
|---|---|
| Message | 160s – 1000s |
| 4x4 | 100s – 1800s |
| X | 25s – 800s |
| Square | 60s – 2600s |

Here, we considered only the times resulting in a completed puzzle. As seen in Fig. 12, the type of information did not significantly affect the shortest times on three out of four of the puzzles but participants receiving conceptual information (the "concept" group) were able to complete the "square" puzzle 3-4 times faster than participants receiving no information or policy information.

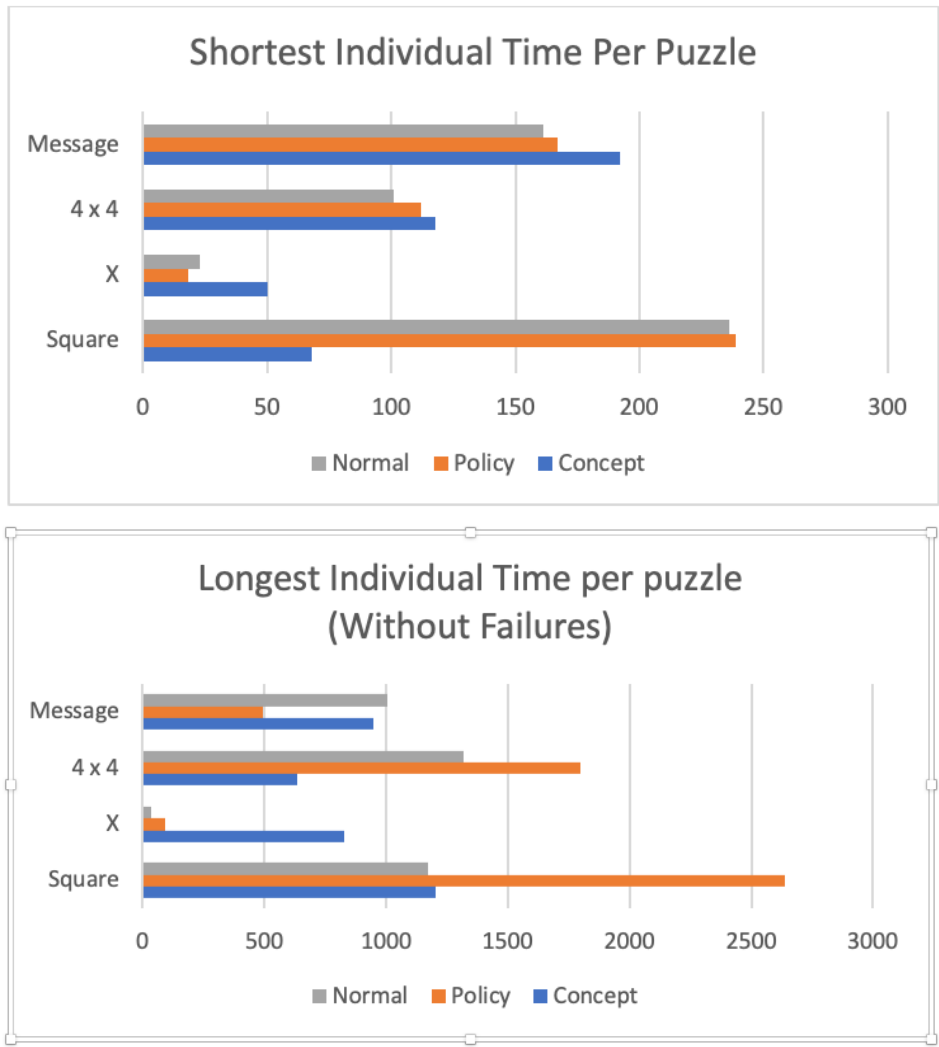

**Fig. 12** Shortest and Longest Time Per Puzzle.

**4.6 Lowest and Highest Individual Number of Attempts per Puzzle**

Participants were allowed to attempt a puzzle multiple times. The number of attempts before achieving a successful completion varied with the "4x4" and the "square" puzzle being the most difficult to solve.

| | |
|---|---|
| Message | 1 – 6 attempts |
| 4x4 | 1 – 14 attempts |
| X | 1 – 6 attempts |
| Square | 1 – 19 attempts |

As seen in Fig. 13, all four puzzles were able to be solved in one or two attempts regardless of the type of information received as a hint. The exception is the "square" puzzle. Without any hint at all (the "normal" group) participants required at least seven attempts to achieve success. However, with some information (the "policy" and "conceptual" groups), participants were able to solve the "square" puzzle in only one or two attempts. The effect of type of information on "square" puzzle performance is also seen when considering the highest number of attempts per puzzle as seen in Fig. 13. Participants receiving no information (the

"normal" group) required as many as 19 attempts to complete whereas participants receiving policy information (the "policy" group) required fewer attempts and participants receiving conceptual information (the "concept" group) required far fewer attempts. The "concept" group required almost one-half the number of attempts as the "normal" group.

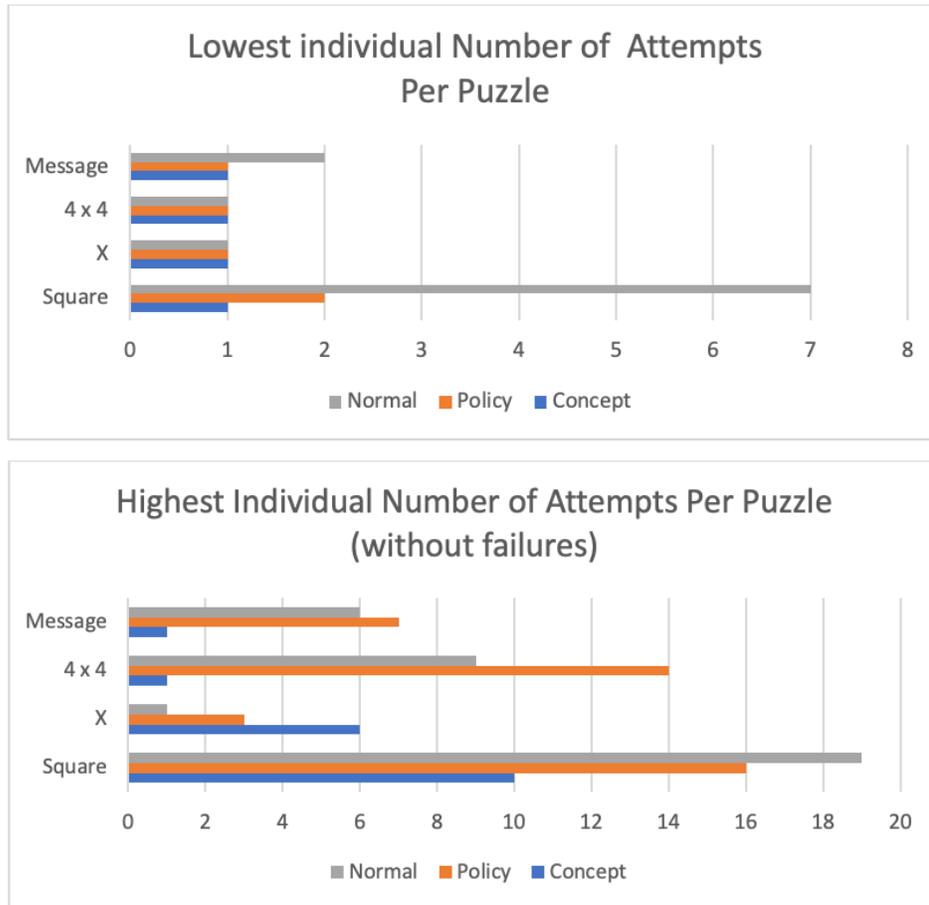

**Fig. 13** Lowest and Highest Individual Number of Attempts Per Puzzle.

## 5   Conclusion

We have confirmed the hypothesis described earlier:

> **H1:** The degree of cognitive augmentation achieved by humans working together on a task in collaboration with cognitive systems is dependent of the nature of information contributed by the cognitive system.

Cognitive performance of the human participants was enhanced to differing degrees when receiving information in the form of a hint. When presented with two different types of information as a hint on how to solve a set of puzzles, *conceptual* information improved performance more than *policy/principle* information. Also, *conceptual* and *policy/principle* information improved human performance over participants receiving no information at all as a hint.

Based on these results, when humans collaborate with cognitive systems as a team, we expect to see a greater degree of cognitive augmentation when the cog provides conceptual information to the human.

Cognitive accuracy was increased by 200% using conceptual information. Cognitive precision was increased by 63%-65% when using conceptual information. Cognitive power was increased by 2.5 times (150) when using conceptual information.

These results should be taken into consideration by cognitive system designers and developers to tailor the way in which the cognitive systems assist their human partners. Careful attention should be given to the nature of information provided to the human by the cog.

## 6    Further Research and Discussion

It is important to note the experimental results reported in this paper use only *conceptual* and *policy/principle* types of information. Further studies should include other types of information identified by [21, 22, 23]: *procedure, process, structure, classification, and fact*. Is there a type of information able to achieve even higher levels of cognitive augmentation than *conceptual*?

It is also important to note the cognitive effort needed to solve the four puzzles in our experiment represent only a fraction of possible cognitive efforts to be examined. Future studies should utilize a vast array of cognitive efforts and seek to use cognitive effort tested and scored in other studies. Has "the type of information leading to different levels of cognitive augmentation" phenomenon been observed in other studies already?

When running similar experiments in the future it would be of value to capture age, gender, and other identifying information. This could lead to discovering if the effects of certain types of information differ for different age groups, gender groups, etc.

We realize the wording and presentation of the information given as hints could have an effect. Future studies could present the same type of information in multiple ways to discover if the way information is presented affects cognitive augmentation.